\documentclass[preprint,12pt]{elsarticle}

\makeatletter
\def\ps@pprintTitle{%
   \let\@oddhead\@empty
   \let\@evenhead\@empty
   \def\@oddfoot{\reset@font\hfil\thepage\hfil}
   \let\@evenfoot\@oddfoot
}
\makeatother
\usepackage[left=2.5cm,right=2.5cm,top=3cm,bottom=3cm]{geometry}

\usepackage{amssymb}

\biboptions{compress}

\usepackage{lineno}
\usepackage{braket}
\usepackage{amsmath}
\usepackage{xcolor}
\usepackage{url}
\usepackage{braket}
\usepackage{bm}
\usepackage{listings}
\usepackage{array}
\newcommand{\mf}[1]{\mathbf{#1}}

\newcommand{\bs}[1]{\boldsymbol{#1}}

\def\irrep{\texttt{IrRep}}

\newcounter{bla}

\DeclareFixedFont{\ttb}{T1}{txtt}{bx}{n}{12} 
\DeclareFixedFont{\ttm}{T1}{txtt}{m}{n}{12}  

\usepackage{listings}
\usepackage{hyperref}
\definecolor{codegreen}{rgb}{0,0.6,0}
\definecolor{codegray}{rgb}{0.5,0.5,0.5}
\definecolor{codepurple}{rgb}{0.58,0,0.82}
\definecolor{backcolour}{rgb}{0.95,0.95,0.92}

\lstdefinestyle{mystyle}{
  backgroundcolor=\color{backcolour},   commentstyle=\color{codegreen},
  keywordstyle=\color{magenta},
  numberstyle=\tiny\color{codegray},
  stringstyle=\color{codepurple},
  basicstyle=\ttfamily\footnotesize,
  breakatwhitespace=false,         
  breaklines=true,                 
  captionpos=b,                    
  keepspaces=true,                 
  numbers=left,                    
  numbersep=5pt,                  
  showspaces=false,                
  showstringspaces=false,
  showtabs=false,                  
  tabsize=2
}

\journal{Computer Physics Communications}

\begin{document}

\begin{frontmatter}



\title{\irrep: symmetry eigenvalues and irreducible representations of \textit{ab~initio} band structures}


\author[a,b]{Mikel~Iraola\corref{mikel}}
\author[b] {Juan~L.~Ma\~nes}
\author[c] {Barry~Bradlyn}
\author[d] {Titus~Neupert}
\author[a,e]{Maia~G.~Vergniory\corref{maia}}
\author[d]{Stepan~S.~Tsirkin
}

\cortext[mikel] {\textit{E-mail:} mikel.i.iraola@gmail.com}
\cortext[maia] {\textit{E-mail:} maiagvergniory@dipc.org}

\address[a]{Donostia International Physics Center, 20018 Donostia-San Sebastian, Spain}
\address[b]{Department of Condensed Matter Physics, University of the Basque Country UPV/EHU, Apartado 644, 48080 Bilbao, Spain}
\address[c]{Department of Physics and Institute of Condensed Matter Theory,University of Illinois at Urbana-Champaign, IL 61801, USA}
\address[d]{Department of Physics, University of Zurich, Winterthurerstrasse 190, CH-8057 Zurich, Switzerland}
\address[e]{IKERBASQUE, Basque Foundation for Science, Maria Diaz de Haro 3, 48013 Bilbao, Spain}

\begin{abstract}

We present \irrep{} -- a Python code that calculates the symmetry eigenvalues of electronic Bloch states in crystalline solids and the irreducible representations under which they transform. As input it receives bandstructures computed with state-of-the-art Density Functional Theory codes such as VASP, Quantum Espresso, or Abinit, as well as any other code that has an interface to  Wannier90. Our code is applicable to materials in any of the 230 space groups and double groups preserving time-reversal symmetry with or without spin-orbit coupling included, for primitive or conventional unit cells. This makes \irrep{} a powerful tool to systematically analyze the connectivity and topological classification of bands, as well as to detect insulators with non-trivial topology, following the  Topological Quantum Chemistry formalism: \irrep{} can generate the input files needed to calculate the (physical) elementary band representations and the symmetry-based indicators using the  \href{https://www.cryst.ehu.es/cgi-bin/cryst/programs/magnetictopo.pl}{\textit{\textcolor{blue}{CheckTopologicalMat}}}  routine of the    Bilbao Crystallographic Server. It is also particularly suitable for interfaces with other plane-waves based codes, due to its  flexible structure. 



\end{abstract}



\end{frontmatter}



{\bf PROGRAM SUMMARY}

\begin{small}
\noindent
{\em Program Title:} \irrep{}      \\
{\em Developer's repository link:} \href{https://github.com/stepan-tsirkin/irrep}{https://github.com/stepan-tsirkin/irrep} \\
{\em Licensing provisions:} GPLv3\\
{\em Programming language:} Python                                   \\
{\em Nature of problem:} Symmetry properties of electronic band structures in solids are tightly related to their topological features. This relation is set mathematically by the formalisms of Topological Quantum Chemistry [1,2] and symmetry-based indicators of topology [3], but their application requires knowledge of the irreducible representations of the bands. Therefore, a code to calculate irreducible representations of \textit{ab initio} bands is essential for a systematic theoretical search and classification of topological materials.\\
{\em Solution method:} \irrep{} reads wave functions from files generated by VASP, Abinit, or Quantum Espresso (also files written as input for Wannier90), determines the space group and symmetry operations via the {\tt spglib} library, evaluates the eigenvalues of the symmetry operations for the selected bands, and applies Group Theory to determine their irreducible representations.\\
{\em Additional comments:} In Abinit calculations, \texttt{istwfk=1} should be used. Routines to get Wannier charge centers and Zak phase are also included, although they work reliably only with norm-conserving pseudopotentials.\\
   \\



\end{small}

\section{Introduction}

Symmetries are fundamental to the properties of quantum systems\cite{weyl1950theory}. 
In particular, knowledge of the symmetry operations under which energy eigenstates transform is crucial to determine the degeneracy of energy levels, the allowed couplings to external fields, and the effect of symmetry-breaking perturbations. For instance, the degeneracies present in the vibrational spectrum of a molecule are determined by the dimensions of its symmetry group's irreducible (co-)representations (IRs). For electrons in periodic crystals,  the band structure of a material in reciprocal space cannot be characterized without studying the symmetry properties of its wave functions\cite{bouckaert1936theory,Bra-Cra}: symmetries can protect or prevent band crossings, predict splittings produced by spin-orbit coupling, and explain gap openings coming from specific terms in the Hamiltonian.

With the discovery of topological insulators\cite{PhysRevLett.95.226801,PhysRevLett.96.106802}, the symmetry analysis of band structures has regained importance. Recently it was discovered that electrons in periodic lattices with crystalline symmetries can yield rich physics due to the interplay of symmetry and topology\cite{FuKane,Teo08,Turner2010,Hughes11,Fu2011,Hsieh2012,Turner2012}. 
Two main developments in the application of symmetry to the identification and classification of topological insulators gave a gigantic push to the field: First, the theory of Topological Quantum Chemistry (TQC)\cite{tqc1,tqc2,tqc3} built upon physical elementary band repesentations (PEBRs) classified all possible atomic limits in all nonmagnetic materials, identifying topological bands as those that cannot be expressed as a sum of band representations. Among the topological bands, a subset can be distinguished from trivial bands by computing a set of symmetry-based indicators\cite{cano2020band,po2020symmetry,elcoro2020application} from the irreducible representations under which the bands transform. The set of symmetry-based indicators in each space group can be computed from the band representations, and have been tabulated in \cite{Song2018,Po2017}.
Calculation of the IRs of bands at high symmetry points is fundamental for the application of these methods and has lead to the prediction of many topological insulators\cite{tqcdatabase,Zhang2019,ashvin-materials,xu2020high} and new phases\cite{Ashvin_fragile,Schindlereaat0346,Hoti_Bi,Schindler2019,MBP_fragile,PhysRevB.99.045140,wieder2018axion,wieder2020strong,shi2019charge,gooth2019evidence,wang2019higher}. 

In this work we present \irrep, a robust, open source Python code that calculates symmetry eigenvalues and IRs of the wave-functions at high symmetry points in reciprocal space for any band structure computed by means of 
 Density Functional Theory (DFT). 
Currently, \irrep{} can interface directly with 3 widely used plane-wave DFT codes: the Vienna Ab initio Simulation Package (VASP)\cite{vasp}, Abinit\cite{Abinit} and Quantum Espresso (QE)\cite{QE}. \irrep{} can also read input files in Wannier90 (W90)\cite{W90} format ({\tt .win}, {\tt .eig}, {\tt \_UNK}), prepared by interfaces like pw2wannier90. This allows \irrep{} to be used with any code that has a W90 interface, such as SIESTA\cite{siesta}. Furthermore, \irrep{} has been structured in a user friendly format allowing the implementation of routines to interface with any other plane-wave based code. 

Although similar codes exist for VASP\cite{irvsp} ({\it vasp2trace}, used to calculate the topological bands in \cite{tqcdatabase,tqcsite,xu2020high} materials database)  and QE\cite{matsugatani2020qeirreps}, \irrep{} is the only code that does not restrict the user to a single DFT program. Moreover, our code follows the same notation as the popular Bilbao Crystallographic Server (BCS)\cite{BCS} to identify the IRs, which avoids confusion coming from the lack of an official standard notation, especially for spin-orbit coupled systems. Tables of IRs are encapsulated within the code package, so that \irrep{} can determine IRs without extra input from the user. The output is written in a form that is compatible with the
\href{https://www.cryst.ehu.es/cgi-bin/cryst/programs/magnetictopo.pl}{\textit{\textcolor{blue}{CheckTopologicalMat}}} tool of the BCS\cite{tqcdatabase}. As additional functionality, \irrep{} can separate bands by eigenvalues of certain symmetry operator and calculates the $\mathbb{Z}_{2}$  and $\mathbb{Z}_{4}$ topological indices of time-reversal symmetric band structures\cite{PhysRevLett.95.226801}. The code evolved from the routines written for Ref.~\cite{TsirkinSouzaVanderbilt} to determine the eigenvalues of screw rotations. At a testing level,  the code was used in Refs.~\cite{Schindler2019,jacs.9b10147,jacs.0c00809,PhysRevMaterials.3.041202,S_nchez_Mart_nez_2019} for topological quantum chemistry analysis, and in Refs.~\cite{Julen-arxiv-2019,Julen-PRR-2020} to analyze the dipole selection rules for optical matrix elements. According to Ref.~\cite{pepy} it has been downloaded more than 7000 times. In section \ref{S:2} we will introduce the basic concepts of group theory underlying the operation of \irrep{}. In section \ref{S:3} we will present the workflow of the code. Finally, section \ref{S:4} will be devoted to several examples, illustrating the capabilities of \irrep{} code for the analysis of symmetry and topology. 

\section{Symmetry Properties of Bands: Irreducible Representations}
\label{S:2}
In this section we give a brief overview of the application of group theory to electronic Bloch states. However, only the minimal information necessary to introduce the notation and explain the functionality  of the code is provided. More details may be found in classic textbooks such as \cite{Bra-Cra,ITA}.

The group $G$ of symmetry operations that leaves a crystal invariant is called the space group of the crystal. In particular, $G$ contains all translations by vectors of the Bravais lattice. The (infinite) group of all translations is generated by 3 primitive basis vectors of the lattice, and forms the (normal) translation subgroup $T$ of the space group $G$. This allows us to write the coset decomposition of $G$ with respect to $T$,

\begin{equation} \label{eq: coset_dec}
    G = T + g_{1}T + g_{2}T + ...+ g_{N}T,
\end{equation}
where $g_{1}, g_{2},..., g_{N}$ are called the coset representatives of the decomposition. Notice that the set of coset representatives is finite and non-unique (two coset representatives differing by a translation characterize the same coset). The number $N$ of cosets in the decomposition Eq.~(\ref{eq: coset_dec}) is equal to the order of the point group $\bar{G}=G/T$ (though the set of coset representatives themselves need not be a point group for non-symmorphic lattices). Using this decomposition, any space group can be expressed in terms of the coset representatives and lattice vectors of the Bravais lattice. 



The symmetries of the space group have a well-defined action on the Hilbert space of electronic states in the crystal. We denote by $U_{g}$ the representation of a certain symmetry operation $g\in G$ on the Hilbert space. Since every $g\in G$ is a symmetry of the crystal, each representation matrix $U_g$ commutes with the  Hamiltonian matrix $H$, i.e., $[U_{g},H]=0$. 
Note that $H$ is block diagonal in reciprocal space and each block $H(\bs{k})$ can be put in correspondence to a vector $\bs{k}$ belonging to the first Brillouin zone (BZ). Although the whole matrix $H$ commutes with $U_{g}$, for a general $g\in G$, a block $H(\bs{k})$ may not commute, but rather must be linearly related to $H(g{\bs k})$,
i.e. the operation $g$ transforms $\bs{k}$ into another reciprocal vector $g\bs{k}$.
If we use  Wigner-Seitz notation $g = \{R|\bs{v}\}$ for the space group operations, then $g\bs{k} = R\bs{k}$. The set of $g\in G$ that leaves $\bs{k}$ invariant (up to a reciprocal lattice vector $\bs{G}$) is called the little group $G_{\bs{k}}$ of $\bs{k}$.

Note that the coset decomposition of Eq.~\eqref{eq: coset_dec} can also be applied to the little group $G_{\bs{k}}$:

\begin{equation}\label{eq: coset_decomp_LG}
G_{\bs{k}} = T + g_{1}^{\bs{k}}\ T + g_{2}^{\bs{k}}\ T + ...+ g_{M}^{\bs{k}}\ T,
\end{equation}
where $M\leq N$ since $G_{\bs{k}} < G$. The coset representatives $g_{i}^{\bs{k}}$ can all be chosen to be point group elements only if $G_{\bs{k}}$ is a symmorphic space group. For example, a group containing screw rotations or glide reflections does not contain its point group as a subgroup. In any case, the rotational parts $R_{i}$ of the representatives $g_{i}^{\bs{k}}=\{R_{i}|\bs{v_{i}}\}$ do form a point group $\bar{G}_{\bs{k}}$, called little co-group of $\bs{k}$. While the little group $G_{\bs{k}}$ is infinite, since it contains all translations, the little co-group $\bar{G}_{\bs{k}}$ is finite.


Consider a set $\{\ket{\Psi_{1\bs{k}}},\ket{\Psi_{2\bs{k}}},...,\ket{\Psi_{D\bs{k}}}\}$ of eigenstates of $H(\bs{k})$, closed under the action of $G_{\bs{k}}$. When a symmetry operation $g\in G_{\bs{k}}$ acts on $\ket{\Psi_{n\bs{k}}}$, the state undergoes a linear transformation

\begin{equation}\label{eq:representation}
g\ket{\Psi_{i\bs{k}}}=\sum_{j=1}^{D} K^{ji}(g)\ket{\Psi_{j\bs{k}}}.
\end{equation}

\noindent The matrices $K(g)$ in Eq.~\eqref{eq:representation} form the representation $K$ of $G_{\bs{k}}$ defined in the invariant space spanned by $\{\ket{\Psi_{n\bs{k}}}\}_{D}$. It is said that $K$ is an IR, if this space cannot be divided into smaller  non-trivial invariant subspaces. Every representation is characterized by the set of traces $\chi_{K}(g)=\mathrm{Tr} K(g)$, known as the \textit{character} of the representation.

In general, the closed set $\{\ket{\Psi_{n\bs{k}}}\}_{D}$ contains eigenstates transforming under more than one IR of $G_{\bs{k}}$, meaning that the whole representation $K$ is reducible and can be decomposed as a combination of these IRs
\begin{equation}\label{eq:decomposition_RR}
K=\oplus_{\alpha} m^{\bs{k}}_{\alpha}K_{\alpha},
\end{equation}
where $K_{\alpha}$ is the $\alpha^{\mathrm{th}}$ IR of $G_{\bs{k}}$. Its multiplicity $m^{\bs{k}}_{\alpha}$ can be computed by means of the following expression, often referred to as the  \textit{magic formula}\cite{BCSDV,Serre}
\begin{equation}\label{eq:magicformula}
m^{\bs{k}}_{\alpha}=\dfrac{1}{||\bar{G}_{\bs{k}}||}\sum_{g\in \{g_i^{\bs{k}}\} } \chi_{K}^{*}(g)\chi_{\alpha}(g),
\end{equation}
with $\{g_i^{\bs{k}}\}$ denoting the set of coset representatives in the decomposition of $G_{\bs{k}}$, and $||\bar{G}_{\bs{k}}||$ the number of symmetry operations in the little cogroup $\bar{G}_{\bs{k}}=G_{\bs{k}}/T$. $\chi_{\alpha}$ and $\chi_{K}$ indicate the characters corresponding to the IR $K_{\alpha}$ and the representation $K$ to be decomposed, respectively. \irrep{} uses this formula to determine the IRs of the eigenstates of $H(\bs{k})$ at  high-symmetry points in the BZ.

\section{Implementation in \irrep{} code}
\label{S:3}

In this section, we present the workflow of the \irrep{} code and describe its main functionalities and the particularities of the interface to each DFT software.

\subsection{Reading DFT data and input parameters}\label{sub:read_data}

To keep the interaction with the user simple, \irrep{} reads as much needed information as possible from the DFT code’s output files. Only parameters determining the user-defined task should be given in the command line in the format 
\begin{lstlisting}[language=bash, breaklines=True]
python -m irrep <keyword1>=<value1> <keyword2>=<value2> ...
\end{lstlisting}
Among all options, the most important ones can be found in Tab.~\ref{T:run_opt}.  Depending on the DFT code used to calculate wave functions, a different interface should be chosen. The interfaces are selected  with the keyword \texttt{code}; currently, it includes interfaces to VASP~\cite{vasp}, Abinit~\cite{Abinit} and QE~\cite{QE}. It can also read the input files for W90~\cite{W90}, which allows  \irrep{} to be used with any of the multiple codes that support the Wannier90 interface. Note that no user-defined input file is needed, and the rest of the needed information will be read from the DFT output files (see Table~\ref{tab:files}).
\begin{table}[t] 
	\centering
	\footnotesize
	\begin{tabular}{l l}
		\hline\hline
		\textbf{Keyword} & \textbf{Function}\\
		\hline
		\texttt{fWAV} & VASP input file with wave functions. Default: WAVECAR \\
		\texttt{fPOS} & VASP input file with the crystal structure. Default: POSCAR  \\
		\texttt{fWFK} & Abinit input file with wave functions \\
		\texttt{prefix} & variable \texttt{prefix} in QE calculation or \texttt{seedname} in W90 \\
		\texttt{IBstart} & first band to be considered \\
		\texttt{IBend} & last band to be considered \\
		\texttt{code} & name of the DFT code: {\tt vasp}, {\tt espresso}, {\tt abinit}, {\tt wannier90}\\
		\texttt{spinor} & whether wave functions are spinors  or not ({\tt T/F}, only needed for VASP) \\
		\texttt{Ecut} & plane wave cutoff to be applied (in eV), recommended $\approx 50$ eV \\
		\texttt{kpoints}  & indices of $k$-points at which IRs must be computed \\
		\texttt{kpnames}  & labels of $k$-points at which IRs must be computed \\
		\texttt{refUC} & transformation of basis vectors with respect to standard setting \\
		\texttt{shiftUC} & shift of origin with respect to standard setting \\
		\texttt{onlysym} & stop after finding symmetries \\
		\texttt{ZAK}  & calculate Zak phases \\
		\texttt{WCC} & calculate wannier charge centers \\
		\hline
	\end{tabular}
	\caption{Principal keywords to fix running options with \irrep{} and their function.}
	\label{T:run_opt}
\end{table}
\begin{table}[bp]
    \centering
    	\footnotesize
    \begin{tabular}{c|c|c}
    interface & {\tt code=} & files \\\hline 
    VASP & {\tt vasp} & {\tt POSCAR} and {\tt WAVECAR}\\ 
    Abinit & {\tt abinit} & {\tt *\_WFK}\\
    QE & {\tt espresso} & {\tt *.save/data\_file\_schema.xml} \\
    & & and \texttt{*.save/wfc*.dat}\\
    Wannier90 & {\tt wannier90} & {\tt *.win}, {\tt *.eig}, {\tt UNK*}
    \end{tabular}
    \caption{Files read by \irrep{} depending on the chosen interface}
    \label{tab:files}
\end{table}
While most of the keywords are self-explanatory, the meaning of the keywords \texttt{refUC} and \texttt{shiftUC} requires some elaboration. The tables of IRs are written for the conventional unit cell corresponding to the space group,
i.e. the cell whose lattice vectors are parallel to the symmetry directions of the lattice.
However, the DFT calculation may be done in any unit cell. Let $\{\bs{a_{1}},\bs{a_{2}},\bs{a_{3}}\}$ denote the basis vectors of the cell adopted for the calculation and $\{\bs{c_{1}},\bs{c_{2}},\bs{c_{3}}\}$ those of the conventional setting, \texttt{refUC} is the $3\times3$ matrix $M$ that expresses the relation between them, according to the following expression
\begin{equation} \label{eq: refUC}
	(\bs{c_{1}},\bs{c_{2}},\bs{c_{3}})^{T}=M(\bs{a_{1}},\bs{a_{2}},\bs{a_{3}})^{T}
\end{equation}
Similarly, \texttt{shiftUC} describes the shift of the origin with respect to the origin of the   conventional unit cell. Note that \texttt{shiftUC} and \texttt{refUC} are relevant only to determine the names of IRs in the notation of BCS. The characters can be computed with any choice of unit cell. 

\begin{figure}[ht]
	\centering
	\includegraphics[width=0.8\linewidth]{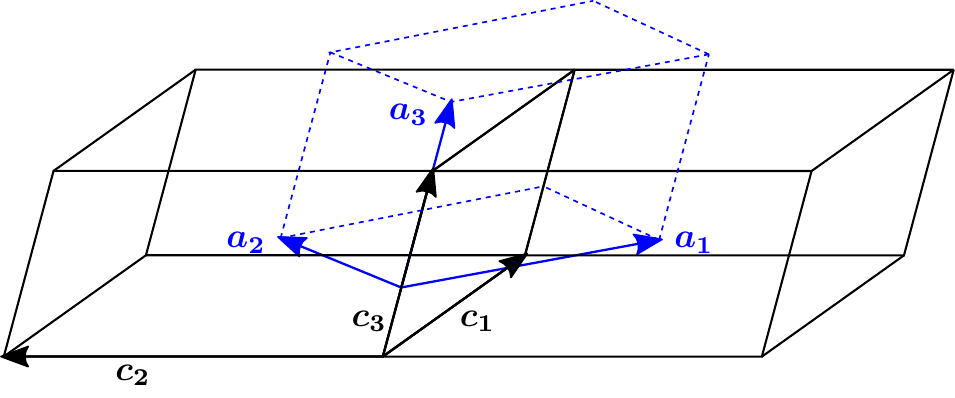}
	\caption{The two  choices considered for the  unit cell of the C-centered monoclinic structure. Conventional (primitive) basis vectors are indicated in black (blue). The conventional unit cell  is marked by solid black lines, while the primitive cell is dashed blue.}
	\label{fig:fig_refUC}
\end{figure}

In order to illustrate the use of these keywords, we work out the example of the C-centered monoclinic structure. Let the relation between the conventional basis vectors and the primitive ones used in the DFT calculation be the following, as  illustrated in Fig.~\ref{fig:fig_refUC}
\begin{displaymath}
\begin{split}
& \bs{c_{1}} = \bs{a_{1}} + \bs{a_{2}}, \\
&\bs{c_{2}} = -\bs{a_{1}} + \bs{a_{2}}, \\
& \bs{c_{3}} = \bs{a_{3}}.
\end{split}
\end{displaymath}
Also, assume that the two origins are related  by the shift $ 0.3\bs{a_{3}}$. Then, the keywords \texttt{refUC} and \texttt{shiftUC} should be used with arguments
\begin{lstlisting}[language=bash, breaklines=True]
refUC=1,1,0,-1,1,0,0,0,1 shiftUC=0,0,0.3
\end{lstlisting}
When a basis transformation is applied, symmetry operations will be printed in both basis. For instance, the following lines illustrate the operations (only identity and inversion) printed by the code in the example of the C-centered crystal:
\begin{lstlisting}[language=bash, breaklines=True]
Space group # 12 has 4 symmetry operations  
 #  1
rotation : |  1   0   0 |     in refUC : |  1   0   0 |
           |  0   1   0 |                |  0   1   0 |
           |  0   0   1 |                |  0   0   1 |
 translation : [   0.000000   0.000000   0.000000 ] 
     in the reference unit cell :
     translation : [   0.000000   0.000000   0.000000 ] 
axis: [0. 0. 1.] ; angle =  0 , inversion : False 
 #  2
rotation : | -1   0   0 |     in refUC : | -1   0   0 |
           |  0  -1   0 |                |  0  -1   0 |
           |  0   0  -1 |                |  0   0  -1 |
 translation : [   0.000000   0.000000   0.400000 ] 
     in the reference unit cell :
     translation : [   0.000000   0.000000   0.000000 ] 
\end{lstlisting}

Once keywords are provided, the class \lstinline{BandStructure} reads the basic information from the output files of the DFT code, such as the plane-wave cutoff, Fermi energy, number of bands, etc. In essence, the lattice vectors, positions of atoms and the energies and wavefunctions of electronic states are read in this way (see Sec. \ref{sub:read_WF} for details). These parameters are found in the files listed in Table~\ref{tab:files}.

When the information is read, it is stored in an object of class \lstinline{Bandstructure} which is independent of the ab initio code used. Thus, if an interface to a new ab initio code is needed, one has to simply implement another constructor for the \lstinline{Bandstructure} class.

\subsection{Determination of the space group}\label{sub:det_SG}


Next, basis vectors and atomic positions are passed to the library Spglib \cite{spglib}, whose routine \lstinline{get_spacegroup} gives the name and number of the space group, while \lstinline{get_symmetry} returns the coset representatives of the space group's decomposition with respect to the translation group (see Sec. \ref{S:2} for details). At this point, if the flag \texttt{onlysym} in Tab.~\ref{T:run_opt} was set, \irrep{} prints the crystal structure and aforementioned coset representatives and then stops. Note that this utility can be useful for VASP even before running the DFT calculation to make sure that the configuration described in POSCAR really matches the assumed space group.

\subsection{Reading wave functions}\label{sub:read_WF}

In VASP, Abinit and QE, eigenstates $\ket{\Psi_{n\bs{k}}}$ of $H(\bs{k})$ are expanded in a basis of plane waves $\ket{\bs{k}+\bs{G}}$:
\begin{equation}\label{eq:expansion_pw}
\ket{\Psi_{n\bs{k}}}=\sum_{\bs{G}}C_{n\bs{k}}(\mf{G})\ket{\bs{k}+\bs{G}},
\end{equation}
where the sum runs over all the reciprocal lattice vectors $\bs{G}$ whose energy is smaller than a cutoff, i.e. $\hbar^2(\bs{k}+\bs{G})^{2}/2m_e<E_{\mathrm{cut}}$. The cutoff coincides with the value indicated by the user if $E_{\mathrm{cut}}$ in Tab.~\ref{T:run_opt} was set; otherwise, it will be the cutoff used in the DFT calculation. After testing the code with different systems, we have noticed that usually a value $E_{\mathrm{cut}}\sim 50$~eV yields accurate results, since the most dominant coefficients in Eq.~\eqref{eq:expansion_pw} correspond to short $\bs{G}$. After the application of the cutoff, the eigenstates $\ket{\Psi_{n\bs{k}}}$ are normalized.

If the DFT calculation ran with PAW pseudopotentials \cite{paw,VASPpaw,AbinitPAW}, the expansion Eq.~\eqref{eq:expansion_pw}  gives the smooth pseudo-wavefunctions $\ket{\tilde{\Psi}_{n\bs{k}}}$, which are related to the all-electron wavefunctions $\ket{\Psi_{n\bs{k}}}$ by a linear transformation $\ket{\Psi_{n\bs{k}}}=\mathcal{T}\ket{\tilde{\Psi}_{n\bs{k}}}$. Note that  $\ket{\tilde{\Psi}_{n\bs{k}}}$ and $\ket{\Psi_{n\bs{k}}}$  transform under symmetry operations in the same way. Hence for simplicity we work with the pseudo-wavefuncions.

In the Wannier90 input files (\texttt{UNK*}) the wavefunctions are written on a real-space grid. In that case we perform a fast Fourier transform (FFT) to obtain the coefficients $C_{n\bs{k}}(\mf{G})$ of Eq.~\eqref{eq:expansion_pw}.

\subsection{Calculation of traces}\label{sub:calc_tr}

For each $\bs{k}$, symmetry operations $g$ of its little group are picked one by one and expectation values $\bra{\Psi_{n\bs{k}}}g\ket{\Psi_{n\bs{k}}}$ are calculated. Note that since the transformation properties of plane waves under translations are trivial, we need only iterate through the coset representatives $g_i^{\bs{k}}$. The calculation of the overlaps depends on whether the DFT calculations were performed on scalar or spinor wavefunctions. Let us consider a $g=\{R|\mf{v}\}\in G_{\bs{k}}$ and show the calculation in both cases:
\begin{itemize}
	\item scalar wavefunctions:
	\begin{equation}\label{eq:tr_wsoc1}
	\bra{\Psi_{n\bs{k}}}g\ket{\Psi_{n\bs{k}}}=\sum_{\bs{G}\bs{G}'} C_{n\bs{k}}^{*}(\bs{G}')C_{n\bs{k}}(\bs{G})\bra{\bs{k}+\bs{G}'}\{R|\mf{v}\}\ket{\bs{k}+\bs{G}}.
	\end{equation}
	From the transformation property of plane-waves,
	\begin{equation}\label{eq:pw_transf}
	g\ket{\bs{k}+\bs{G}}=e^{-i(R\bs{k}+R\bs{G})\cdot\mf{v}}\ket{R\bs{k}+R\bs{G}}, 
	\end{equation}
	together with their orthogonality property,
	\begin{equation}\label{eq:ortho_pw}
	\braket{{\bs{k}'+\bs{G}'}|{\bs{k}+\bs{G}}}=\delta_{\mf{G'},\bs{k}-\bs{k}'+\bs{G}},
	\end{equation}
	it follows that Eq.~\eqref{eq:tr_wsoc1} is reduced to:
	\begin{equation}\label{eq:tr_wsoc2}
	\bra{\Psi_{n\bs{k}}}g\ket{\Psi_{n\bs{k}}}=\sum_{G}C_{n\bs{k}}^{*}(R\bs{k}-\bs{k}+R\bs{G})C_{n\bs{k}}(\bs{G})e^{-i(R\bs{k}+R\bs{G})\cdot\mf{v}}.
	\end{equation} 
	
	\item for spinor wavefunctions  $\ket{\Psi_{n\bs{k}}}=(\ket{\Psi_{n\bs{k}}^{\uparrow}} \ket{\Psi_{n\bs{k}}^{\downarrow}})^{T}$ the matrix element reads:
	\begin{equation}\label{eq:spinor_tr}
	\bra{\Psi_{n\bs{k}}}g\ket{\Psi_{n\bs{k}}}=\sum_{\sigma\sigma'}S_{\sigma\sigma'}(g)\bra{\Psi_{n\bs{k}}^{\sigma}}g\ket{\Psi_{n\bs{k}}^{\sigma'}},
	\end{equation}
	where brakets $\bra{\Psi_{n\bs{k}}^{\sigma}}g\ket{\Psi_{n\bs{k}}^{\sigma'}}$ are computed by means of Eq.~\eqref{eq:tr_wsoc2} and $S(g)$ is an SU(2) matrix corresponding to $g$.
\end{itemize}

After this calculation, \irrep{} adds the expectation values of degenerate eigenstates. Each of these sums is the trace $\chi(g)$ of a matrix $K(g)$ belonging to the representation $K$ defined in the subspace of degenerate eigenstates. These traces might be of interest by themselves, but can also be further used to identify the IRs.

\subsection{Identification of irreducible representations}

The characters $\chi_{\alpha}$ of every IR $K_{\alpha}$ of $G_{\bs{k}}$ were obtained from the BCS~\cite{BCSDV} and are provided with the \irrep{} module. For each subspace of degenerate eigenstates, the magic formula \eqref{eq:magicformula} is applied, yielding the multiplicity $m_{\alpha}^{\bs{k}}$ of IR $K_{\alpha}$ in the subspace of degenerate states. Notice that this procedure detects accidental degeneracies, which happen when eigenstates transforming under different IRs have the same energy.

At this point, the IR of each set of eigenstates is printed, together with the character of the IR. Furthermore, \irrep{} also writes a file \texttt{trace.txt}, which can be passed directly to the program \href{https://www.cryst.ehu.es/cgi-bin/cryst/programs/magnetictopo.pl}{\textit{\textcolor{blue}{CheckTopologicalMat}}} of BCS \cite{BCS}, in order to get information about (physical) elementary band representations and symmetry-based indicators \cite{tqc1,Song2018,Po2017,tqcdatabase} to diagnose the band topology.

By default, the procedure is performed for all the bands calculated by the DFT code. Nevertheless, the user can set values for \texttt{IBstart} and \texttt{IBend} in order to consider only bands in the range [\texttt{IBstart},\texttt{IBend}]. This can be used to noticably shorten the calculation time. Moreover, for the selected set of bands, the smallest direct and indirect gaps with respect to higher bands will be printed. For centrosymmetric crystals, the number of inversion-eigenvalue inversions (band inversions) and the $\mathbb{Z}_{2}$ index \cite{FuKane,MooreZ2}  will also be printed. In Sec.~\ref{s:example_cubi2o4}, we show an example of the output generated by \irrep{} where all these features are represented.

\subsection{Separation by symmetry eigenvalues}

When writing the IRs, \irrep{} can separate the states by their eigenvalues with respect to a certain symmetry operation, if the index of that symmetry was given as \texttt{isymsep} (Tab.~\ref{T:run_opt}). Moreover, wave functions can be grouped by Kramers pairs, via the key \texttt{groupKramers}. The energies will also be written in a file, which can be used to plot bands. This data will be useful if the DFT calculation was done for an ordered set of $k$-points following a certain path in the BZ. 
Bands  with different symmetry eigenvalues will be written in different files. The purpose of this feature is to respect the separation in the plot of bands, which is useful to study the role of symmetries in the protection of band crossings \cite{TsirkinSouzaVanderbilt}.

\irrep{} also contains routines to calculate the Zak phase and Wannier charge centers of a given set of bands (see keys ZAK and WCC in Tab.~\ref{T:run_opt}). 

Note that these functionalities work stably only for calculations employing norm-conserving pseudopotentials, and \texttt{Ecut} should not be specified in the command line (thus the DFT cutoff will be used). With the PAW method, due to the lack of consideration of the all-electron wavefunction, the results for symmetry separation and ZAK phase may be unreliable.

\section{Example materials} 
\label{S:4}

In this section, we present the application of \irrep{} to two material examples, with different symmetries and topology, and analyzed by different DFT codes. With these examples, we cover the main functionalities of the code and also the subtlety related to the transformation between primitive and conventional basis.

\subsection{Irreducible representations in CuBi$_{2}$O$_{4}$} \label{s:example_cubi2o4}

\begin{figure}[t]
	\centering
	\includegraphics[width=01\linewidth]{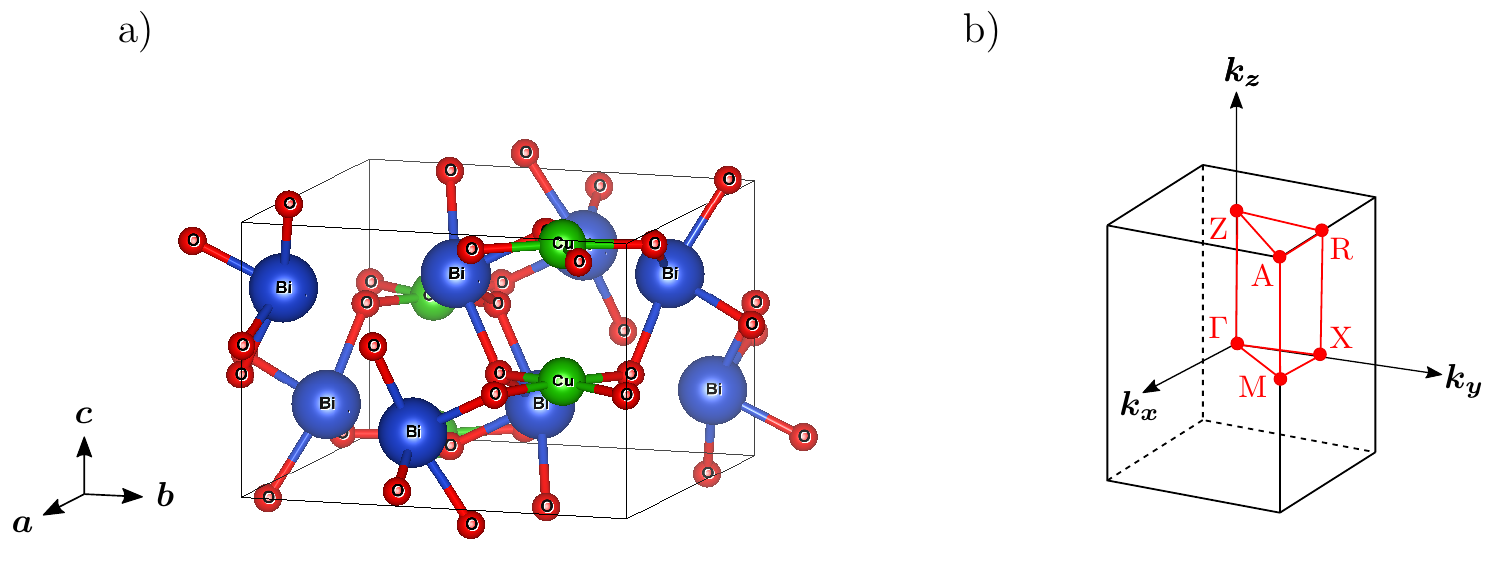}
	\caption{Crystal structure of CuBi$_{2}$O$_{4}$. (a) Unit cell, with Cu, Bi and O atoms in green, blue and red, respectively. (b) BZ and irreducible BZ (red), with high-symmetry points.}
	\label{fig:structurecubi2o4pdf}
\end{figure}

First, we show the application of \irrep{} to CuBi$_{2}$O$_{4}$. In the paramagnetic phase, this material crystallizes in a tetragonal structure characterized by the non-symmorphic space group P4/ncc (No. 130) \cite{Yamada91,Boivin}. Its crystal structure and BZ  are shown in Fig.~\ref{fig:structurecubi2o4pdf}.

An interesting aspect of CuBi$_{2}$O$_{4}$ is found in reciprocal space: the little group of point A=$(1/2,1/2,1/2)$, in the corner of the BZ, has only one (double-valued) IR with dimension 8. Its unusually large dimension makes it promising for the realization of high-degeneracy unconvenctional fermions \cite{Bradlyn16,wieder16}.  Also due to the fact that the number of electrons in the unit cell is not a multiple of 8, this IR forces CuBi$_{2}$O$_{4}$ to be a filling-enforced semimetal. 

\begin{figure}[t]
	\centering
	\includegraphics[width=1.\linewidth]{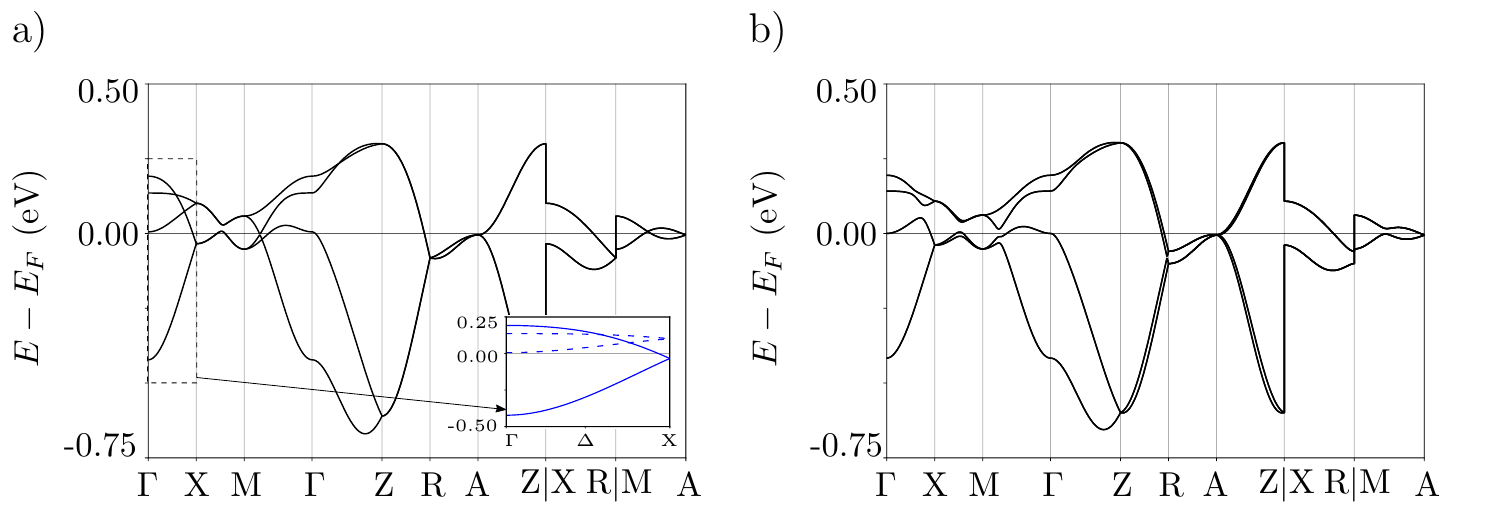}
	\caption{Band structure of CuBi$_{2}$O$_{4}$ (a) without SOC included. Inset: bands in line $\Delta$, connecting $\Gamma$ to X, separated according to their  eigenvalue of symmetry $\{m_{x}|1/2,0,1/2\}$ in the little group: solid (dashed) corresponds to eigenvalue $-1$ ($+1$). (b) Band structure of CuBi$_{2}$O$_{4}$ with SOC included.}
	\label{fig:bandscubi2o4}
\end{figure}

We have calculated the band structure of CuBi$_{2}$O$_{4}$ with Abinit, both treating spin trivially (scalar calculation) and including spin-orbit corrections (spinor calculation). A plane-wave cutoff of 500 eV and cold smearing \cite{coldsmear} were used in the calculation. The BZ was sampled with a grid of $5\times5\times7$. Lattice parameters and atomic positions were taken from the Topological Quantum Chemistry database of materials \cite{tqcdatabase}. The exchange-correlation term was approximated through General Gradient Approximation, in the Perdew Burke Ernzerhof \cite{PBE} parametrization and PAW pseudopotentials were taken from Pseudo Dojo database \cite{pseudodojo}. Computed band structures can be seen in Fig.~\ref{fig:bandscubi2o4}. Output files of \irrep{} are available in the \texttt{examples} folder of \irrep's official Github repository \cite{Irrep_repo}. 

In the rest of the analysis, we focus on the partially-filled isolated set of bands cut by the Fermi level. IRs of the wave functions at high-symmetry points can be calculated by running the following lines (case without SOC):
\begin{lstlisting}[language=bash, breaklines=True]
python -m irrep code=abinit kpnames=GM,X,M,Z,R,A 
	Ecut=100 spinor=F fWFK=CuBi2O4-scalar_WFK 
	IBstart=145 IBend=148
\end{lstlisting}
IRs obtained in this way are written in Tab.~\ref{T:cubi2o4_irreps}. The following lines illustrate part of the output for the point R$=(0,1/2,1/2)$. Even though the little group of R contains many coset representatives, only two of them (represented by indices 1 and 7) are shown here for brevity.

\begin{table}[ht]
	\begin{tabular}{c c c c c c}
		\hline\hline
		$\Gamma$ & X & M & Z & R & A\\
		\hline
		$\Gamma_{4}^{+} \oplus \Gamma_{2}^{+} \oplus \Gamma_{4}^{-} \oplus \Gamma_{2}^{-}$ & X$_{2} \oplus $X$_{1}$ & M$_{4} \oplus $M$_{3}$ & 2Z$_{1}$ & R$_{1}$R$_{2}$ & A$_{3}$A$_{4}$ \\
		$2\bar{\Gamma}_{6} \oplus 2\bar{\Gamma}_{8}$ & $2\bar{\mathrm{X}}_{3}\bar{\mathrm{X}}_{4}$ & $2\bar{\mathrm{M}}_{5}$ & 2$\bar{\mathrm{Z}}_{5}\bar{\mathrm{Z}}_{7}$ & $\bar{\mathrm{R}}_{4}\bar{\mathrm{R}}_{4}\oplus \bar{\mathrm{R}}_{3}\bar{\mathrm{R}}_{3}$ & $\bar{\mathrm{A}}_{5}\bar{\mathrm{A}}_{5}$ \\ 
		\hline\hline
	\end{tabular}
	\caption{IRs at high-symmetry points of the partially filled set of bands of CuBi$_{2}$O$_{4}$. In the second (third) row, IRs of the calculation without (with) SOC included are listed.}
	\label{T:cubi2o4_irreps}
\end{table}


\begin{lstlisting}[language=bash, breaklines=True]
k-point   5 :[0.  0.5 0.5] 
 number of irreps = 8
   Energy  | multiplicity |  irreps  | sym. operations  
           |              |          |     1          7
  -0.1006  |        4     | -R4(2.0) |  4.0+0.0j  -4.0+0.0j
  -0.0595  |        4     | -R3(2.0) |  4.0+0.0j   4.0+0.0j
inversion is # 9
number of inversions-odd Kramers pairs :  2
Gap with upper bands :  2.02
\end{lstlisting}

Interesting information about the bands and even about the chemistry of the system can be extracted from the knowledge of IRs. This set of IRs is consistent with an elementary band representation \cite{tqc1} induced from Wannier functions sitting in Wyckoff position 4c: $(B\uparrow G)_{4c}$ in the case without SOC, $(^{1}\bar{\mathrm{E}}_{2} ^{2}\bar{\mathrm{E}}_{2}\uparrow G)_{4c}$ with SOC, in the notation of Ref.~\cite{tqc1}. According to the framework of band representations \cite{Zak1,Zak2}, which explains how bands in reciprocal space inherit their symmetry properties from orbitals in real space, the Wannier functions that induce these bands transform as a combination of d$_{x^{2}-y^{2}}$ and d$_{xy}$ orbitals.

At every point $\bs{k}$ belonging to the line $\Delta$ that connects $\Gamma$ to X, the little group contains the glide symmetry $g_{x}=\{m_{x}|1/2,0,1/2\}$. This means that for $\bs{k}\in \Delta$, wave functions of bands are also eigenstates of $g_{x}$, so that we can distinguish them by their eigenvalue under this symmetry. As we mentioned, \irrep{} can extract this eigenvalue, by running the option \texttt{isymsep}$=13$, which corresponds to $g_{x}$:

\begin{lstlisting}[language=bash, breaklines=True]
python -m irrep code=abinit Ecut=100 spinor=F 
	fWFK=CuBi2O4-scalar_WFK IBstart=145 IBend=148
	isymsep=13
\end{lstlisting}

The index corresponding to $g_{x}$ can be derived beforehand by running the option \texttt{onlysym}. The result is shown in the inset of Fig.~\ref{fig:bandscubi2o4}(a), where bands with eigenvalue $-1$ (+1) of $g_{x}$ are indicated in solid (dashed). This calculation tells us that the crossings between dashed and solid bands are protected by $g_{x}$ and thus cannot be gapped out without breaking this symmetry. Such criteria can be used to systematically study  symmetry protected band crossings~\cite{TsirkinSouzaVanderbilt}. 

\subsection{Bismuth: high order topological insulator}

In this example, we will present the calculation of $\mathbb{Z}_{2}$ and $\mathbb{Z}_{4}$ indices with \irrep. For that, we will work with a particularly interesting and well-known material: bismuth.

\begin{figure}[t]
	\centering
	\includegraphics[width=0.9\linewidth]{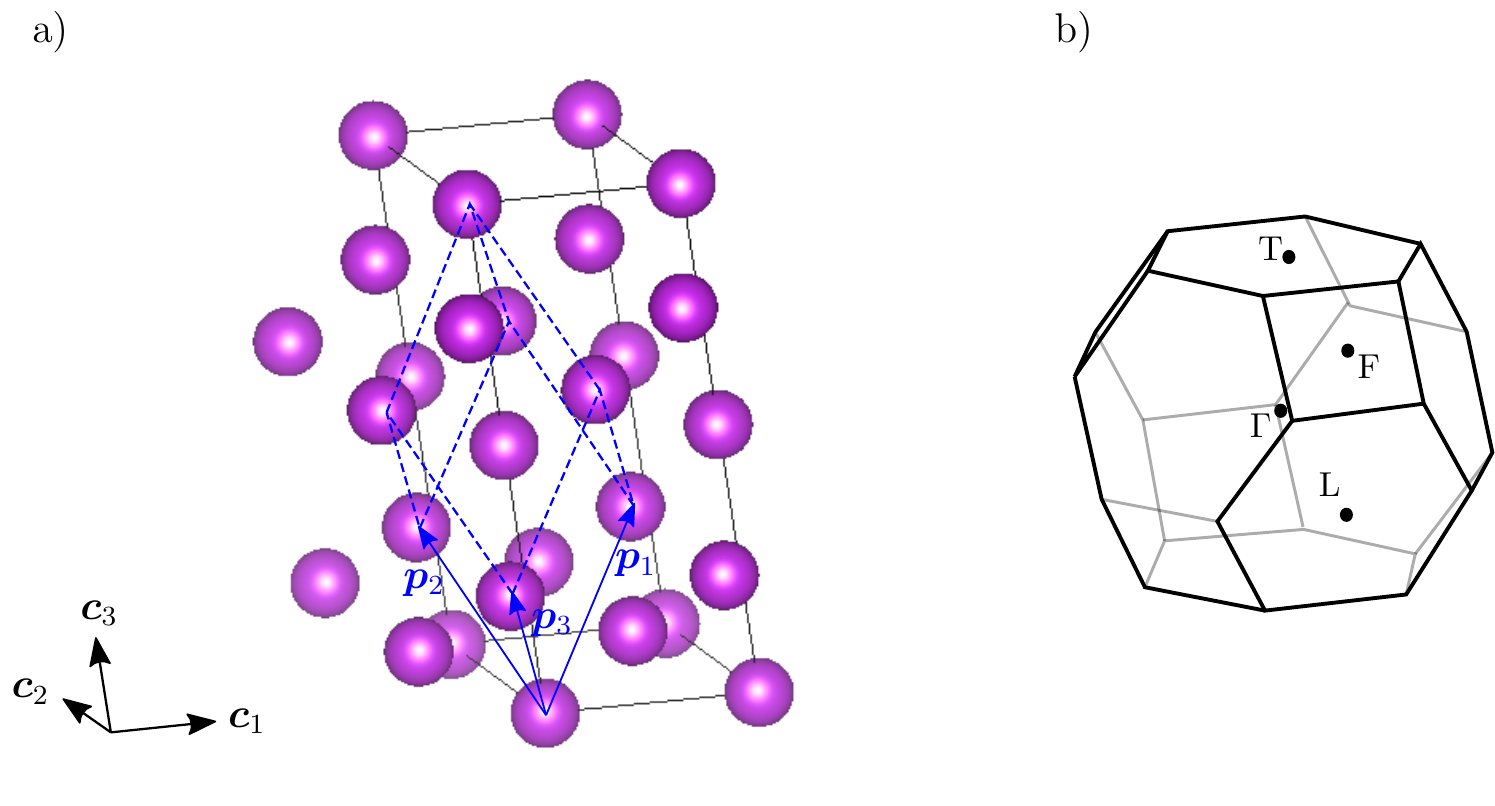}
	\caption{a) Crystal structure of Bi in space group R$\bar{3}m$. Black lines delimit the convenctional unit cell, whose basis vectors are $\{\bs{c}_{i}\}_{i=1,3}$, while blue lines delimit the primitive unit cell used in the DFT calculation. b) Brillouin Zone corresponding to the primitive cell and TRIM (one from each star of $k$-points).}
	\label{fig:cell_BZ_Bi}
\end{figure}

In the presence of only time-reversal symmetry (TRS), an insulator can belong to either the trivial or the topological phase. The system cannot undergo a transition from one phase to the other if the gap is not closed or TRS is not broken in the process. In this spirit, the topology of the system can be characterized by a $\mathbb{Z}_{2}$ invariant \cite{PhysRevLett.95.226801,kane2005z,MooreZ2}, which is -1 (+1) in the topological (trivial) phase. With inversion, the $\mathbb{Z}_{2}$ invariant can be calculated by multiplying the inversion eigenvalues of Kramers pairs of occupied bands at all time-reversal invariant momenta (TRIM)\cite{FuKane}. In the topological case, we say that the system has a band-inversion.

Crystal symmetries may enrich the topology of time-reversal invariant insulators, giving access to new phases, some of which can not be detected by the $\mathbb{Z}_{2}$ index. This is the case for bismuth in space group R$\bar{3}m$: the $\mathbb{Z}_{2}$ index has value +1, which means that the ground state corresponding to the occupied bands in Bi is a trivial insulator as per its $\mathbb{Z}_{2}$ index, according to the discussion above. However, in Ref.~\cite{Hoti_Bi} it was shown that the ground state belongs to a higher-order topological phase, characterized by a $\mathbb{Z}_4$ index equal to 2. Here, we will reproduce with \irrep{} this analysis.

\begin{figure}[t]
	\centering
	\includegraphics[width=1.\linewidth]{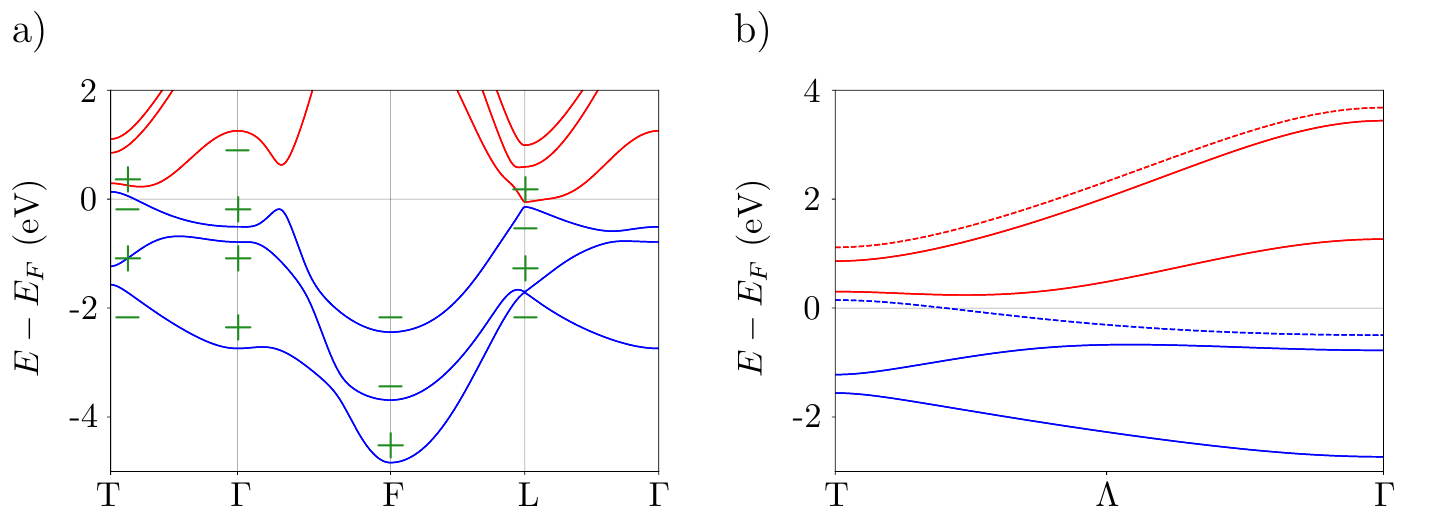}
	\caption{Band structure of Bi in space group R$\bar{3}m$. Since blue and red bands do not touch, we will ignore the electron-hole pockets and speak of occupied (blue) and unoccupied (red) bands. (a) Inversion eigenvalues at TRIM are indicated in green. (b) Bands along C$_{3z}$-invariant line $\Lambda$, which connects T to $\Gamma$; solid and dashed bands have $C_{3z}$ eigenvalues -1 and exp$(\pm i\pi/3)$, respectively.}
	\label{fig:bands_Bi}
\end{figure}

We have used VASP to perform \textit{ab initio} calculations of Bi in the primitive unit cell. All calculations include spin-orbit corrections. A cutoff of 520~eV was set for the plane-wave basis, together with a Gaussian smearing. The BZ was sampled with a grid of $7\times7\times7$ $k$-points. We used PBE prescription as an approximation for the exchange-correlation term and PAW pseudopotentials \cite{PBE}. The calculated bands are shown in Fig.~\ref{fig:bands_Bi}(a). In Fig~\ref{fig:bands_Bi}(b), we show the bands separated by eigenvalues of $C_{3z}$ using \irrep's option \texttt{isymsep}.

Space group R$\bar{3}m$ (No. 166) belongs to the rhombohedral family, in which conventional and primitive unit cells do not match, as can be seen in Fig.~\ref{fig:cell_BZ_Bi}(a). Consequently, in order to use \irrep{} to get the IRs, we need to provide it with the transformation from the primitive to the conventional unit cell, via the keywords \texttt{refUC} and \texttt{shiftUC}. Since the origin of both, primitive and conventional cells is located in the same point, there is no need to specify \texttt{shiftUC}. Equation~\eqref{eq: refUC} takes the following form:
\begin{equation}
    (\bs{c_{1}},\bs{c_{2}},\bs{c_{3}})^{T}=
    \begin{pmatrix}
    1 & -1 & 0 \\
    0 & 1 & -1 \\
    1 & 1 & 1 \\
    \end{pmatrix}
    (\bs{a_{1}},\bs{a_{2}},\bs{a_{3}})^{T},
\end{equation}
so \texttt{refUC}$=1,-1,0,0,1,-1,1,1,1$. In the following lines, which have been taken from the output of IrRep available among the examples in \irrep's Github repository\cite{Irrep_repo}, we show how \irrep{} prints the matrix of symmetry operations, in particular for the 3-fold rotation, in the settings before and after applying the transformation of the unit cell:

\begin{lstlisting}[language=bash, breaklines=True]
 #  3
rotation : |  0   0   1 |     in refUC : |  0  -1   0 |
           |  1   0   0 |                |  1  -1   0 |
           |  0   1   0 |                |  0   0   1 |
spinor   : | 0.500-0.866j -0.000-0.000j |
           | 0.000-0.000j  0.500+0.866j |
 translation : [   0.000000   0.000000   0.000000 ] 
     in the reference unit cell :
     translation : [   0.000000   0.000000   0.000000 ] 
axis: [0. 0. 1.] ; angle = 2/3 pi, inversion : False 
\end{lstlisting}

To make sure that the transformation is correct, one has to check whether matrices and translation vectors after the change of basis match with those in the table file of the corresponding space group. Alternatively, they can be compared to matrices and translations in the \href{https://www.cryst.ehu.es/cryst/get_gen.html}{GENPOS} application of the BCS\cite{GENPOS}. The next step is to calculate the IRs of occupied bands (Tab.~\ref{T:Bi_irreps}). For that, we call \irrep{} with the keywords written in the following lines:

\begin{lstlisting}[language=bash, breaklines=True]
python -m spinor=T code=vasp kpnames=T,GM,F,L Ecut=50 refUC=1,-1,0,0,1,-1,1,1,1 EF=5.2156 IBstart=5 IBend=10
\end{lstlisting}

\begin{table}[ht]
	\begin{tabular}{c c c c}
		\hline\hline
		T & $\Gamma$ & F & L \\
		\hline
		$\bar{\mathrm{T}}_{9} \oplus \bar{\mathrm{T}}_{8} \oplus \bar{\mathrm{T}}_{6}\bar{\mathrm{T}}_{7}$ & $2\bar{\Gamma}_{8} \oplus \bar{\Gamma}_{4}\bar{\Gamma}_{5}$ & $\bar{\mathrm{F}}_{3}\bar{\mathrm{F}}_{4} \oplus \bar{\mathrm{F}}_{5}\bar{\mathrm{F}}_{6} \oplus \bar{\mathrm{F}}_{5}\bar{\mathrm{F}}_{6}$ & $\bar{\mathrm{L}}_{5}\bar{\mathrm{L}}_{6} \oplus \bar{\mathrm{L}}_{3}\bar{\mathrm{L}}_{4} \oplus \bar{\mathrm{L}}_{5}\bar{\mathrm{L}}_{6}$\\
		\hline\hline
	\end{tabular}
	\caption{IRs at TRIM for the last 6 occupied bands in Bi, calculated with \irrep. In each $k$-point, IRs are written from left to right in ascending energy order, e.g., $\bar{\mathrm{T}}_{6}\bar{\mathrm{T}}_{7}$ is higher in energy than $\bar{\mathrm{T}}_{8}$.}
	\label{T:Bi_irreps}
\end{table}

With the knowledge of the IRs and their inversion eigenvalues [see Fig.~\ref{fig:bands_Bi}(a)], we conclude that the total number of (Kramers pairs of) $-1$ inversion eigenvalues in the occupied bands is even, thus the $\mathbb{Z}_2$ index $z_{2}=+1$. However, there are two band inversions between $\Gamma$ and T that the $\mathbb{Z}_{2}$ invariant cannot detect. Indeed, this double band-inversion leads to the $\mathbb{Z}_4$ invariant $z_4=2$, since the number of Kramers pairs of $-1$ inversion eigenvalues is equal to $2 \mod 4$. This means that Bi is a higher-order topological insulator (HOTI) \cite{Hoti_Bi}. The value of the $\mathbb{Z}_{4}$ index and number of $-1$ inversion eigenvalues are, by default, calculated and printed by \irrep; in the following line, we show the way in which they are printed by the code, together with information about the direct and general gaps:\footnote{here the gap refers only to the high-symmetry points included in the calculation. The real gap may be smaller (and even may close) at some arbitrary point away from high-symmetry points. }

\begin{lstlisting}[language=bash, breaklines=True]
Number of inversions-odd Kramers pairs IN THE LISTED KPOINTS: 6   Z4=  2
Minimal direct gap: 0.08857033154551353  eV
Indirect  gap: -0.1886089499035215  eV
\end{lstlisting}

\section{Conclusion}\label{s:conclusion}

\irrep{} is a Python code for the calculation of irreducible representations of DFT calculated bands at high-symmetry points. It is a powerful tool for the detection and classification of topological sets of bands and materials, applicable with calculations performed both with or without SOC and using unit cells that might be non-conventional. Its structure keeps  the implementation of interfaces to plane-wave DFT codes simple; currently, it is compatible with VASP, Abinit, Quantum Espresso and any code that has an interface to Wannier90 (which covers most of the popular DFT codes). Additionally, routines for separating  bands based on an eigenvalue of certain symmetry operation are included. \irrep{} can be freely downloaded from \url{https://github.com/stepan-tsirkin/irrep} and/or installed with \texttt{pip}; the repository also contains examples, including the analysis of CuBi$_{2}$O$_{4}$ that we have presented in this work to illustrate the utility of the code.

\section*{Software availability}
All software used and developed in this article  (except VASP) is open-source and available for free. \irrep{} is available via {\tt pip} \cite{pypi-irrep} and GitHub \cite{Irrep_repo}. 
External libraries used in \irrep{} include {\tt spglib} \cite{spglib}, {\tt NumPy} \cite{numpy}, {\tt SciPy} \cite{scipy}, and  {\tt lazy-propery} \cite{lazy-property}.
VASP is commercial software available from the developers for a fee \cite{vasp-at}. 
Other codes are available at \cite{wannier.org} (Wannier90), \cite{qe.org} (QuantumEspresso) and \cite{Abinit.org} (Abinit). Figures of crystal structures were generated with VESTA \cite{vesta}. Band structures were plotted with BEplot \cite{BEplot_repo}, which uses \texttt{Matplotlib} \cite{matplotlib}. Inkscape vector graphics editor \cite{Inkscape} was used in most figures. 

\section*{Acknowledgements}
M.G.V. and M.I. acknowledges the Spanish Ministerio de Ciencia e Innovacion (grant number PID2019-109905GB-C21) and DFG INCIEN2019-000356 from Gipuzkoako Foru Aldundia. B. B. acknowledges the support of the Alfred P. Sloan foundation, and the National Science Foundation under grant DMR-1945058. The work of J.L.M. has been supported by Spanish Science Ministry grant PGC2018-094626-BC21
(MCIU/AEI/FEDER, EU) and Basque Government grant IT979-16. S.S.T. and T.N. acknowledge support from NCRR Marvel and from the European Union’s Horizon 2020 research and innovation program (ERC-StG-Neupert-757867-PARATOP). S.S.T. also acknowledges support from the Swiss National Science Foundation (grant number: PP00P2\_176877).





\bibliographystyle{elsarticle-num}

\bibliography{biblio}







\end{document}